\documentclass[twocolumn]{jpsj2}

\usepackage{times}
\usepackage{graphicx}

\title{Multipole as $\mib{f}$-Electron Spin-Charge Density
in Filled Skutterudites}

\author{Takashi {\sc Hotta}\thanks{Present address:
Department of Physics, Tokyo Metropolitan University,
1-1 Minami-Osawa, Hachioji, Tokyo 192-0397.}
}

\inst{Advanced Science Research Center,
Japan Atomic Energy Agency, Tokai, Ibaraki 319-1195, Japan}

\recdate{\today}

\abst{
It is shown that $f$-electron multipole is naturally defined as
spin-charge one-electron density operator
in the second-quantized form with the use of tensor operator
on the analogy of multipole expansion of electromagnetic potential
from charge distribution in electromagnetism.
Due to this definition of multipole,
it is possible to determine multipole state
from a microscopic viewpoint on the basis of
the standard linear response theory
for multipole susceptibility.
In order to discuss multipole properties of filled skutterudites,
we analyze a seven-orbital impurity Anderson model
by employing a numerical renormalization group method.
We show our results on possible multipole states of
filled skutterudite compounds.
}

\kword{multipole, spin-charge density, filled skutterudites,
numerical renormalization group method}

\begin{document}
\maketitle

%
%
\section{Introduction}

One of current research trends in material science is
to synthesize new compounds which exhibit exotic phenomena
concerning magnetism and superconductivity.
Due to the improvement of single crystal quality with the use of
modern crystal growth techniques, essentially new phenomena
have been also discovered even in known compounds.
In particular, ordering of higher-rank multipole has been
actively studied both from experimental and theoretical sides
in the research field of strongly correlated $f$-electron systems.
\cite{Kuramoto,Kusunose}
In general, owing to the strong spin-orbit coupling of $f$ electrons,
spin-orbital complex degree of freedom, i.e., multipole, is
considered to be active in $f$-electron compounds.
However, when orbital degeneracy is lifted, for instance,
due to the crystal structure with low symmetry,
only spin degree of freedom often remains.
Namely, $f$-electron compounds crystallizing in the cubic structure
with high symmetry are quite important for the research
of multipole phenomena.

In this sense, filled skutterudite compounds LnT$_4$X$_{12}$
with lanthanide Ln, transition metal atom T, and pnictogen X
provide us an ideal stage for the promotion of multipole physics,
since this material group crystallizes in the cubic structure
of $T_{\rm h}$ point group.\cite{Takegahara}
Moreover, it is possible to synthesize many isostructural materials
with different kinds of rare-earth and actinide ions,
leading to the development of systematic research
on multipole ordering.
In fact, recent experiments in cooperation with
phenomenological theory have revealed that
multipole ordering frequently appears in filled skutterudites.
For instance, a rich phase diagram of PrOs$_4$Sb$_{12}$ with
field-induced quadrupole order has been unveiled
experimentally and theoretically.\cite{Aoki,Tayama,Shiina}
Recently, antiferro $\Gamma_1$-type higher multipole order \cite{Kuramoto}
has been discussed for PrRu$_4$P$_{12}$ \cite{Takimoto,Iwasa}
and PrFe$_4$P$_{12}$.\cite{Kiss,Sakai,Kikuchi}
Except for Pr-based filled skutterudites,
signs of multipole phenomena have been also found.
In NdFe$_4$P$_{12}$, a significant role of quadrupole
at low temperatures has been suggested from the measurement of
elastic constant.\cite{Nakanishi}
In SmRu$_4$P$_{12}$, a possibility of octupole order has been
proposed from several kinds of experiments.
\cite{Yoshizawa,Hachitani,Masaki1,Masaki2,Aoki-Sm-new}

As mentioned above, theoretical research on multipole order
has been developed mainly from a phenomenological viewpoint
on the basis of an $LS$ coupling scheme
for multi-$f$-electron state.
Such investigations could explain several experimental results
and research activity along this direction will
be still important in future.
However, we strongly believe that it is also important
to promote microscopic approach for understanding of
multipole phenomena in parallel with phenomenological research.
Based on this belief, the present author and collaborators
have made effort to develop a microscopic theory
for multipole-related phenomena by exploiting 
a $j$-$j$ coupling scheme.\cite{Hotta1}
In particular, octupole ordering in NpO$_2$ has been clarified
by evaluating multipole interaction with the use of
the standard perturbation method
in terms of electron hopping.\cite{Kubo1,Kubo2,Kubo3}
We have also discussed possible multipole states of
filled skutterudites by analyzing multipole susceptibility of
a multiorbital Anderson model based on the $j$-$j$ coupling scheme.
\cite{Hotta2,Hotta3,Hotta4,Hotta5,Hotta6,Hotta7}

However, it seems to be still difficult to understand intuitively
the physical meaning of multipole degree of freedom
due to the mathematically complicated form of
multipole operator defined by using total angular momentum.
As mentioned above, multipole is considered to be
spin-orbital combined degree of freedom.
In this sense, it seems to be natural to regard multipole
as anisotropic spin and/or charge density.
This point has been first emphasized in the visualization of
octupole order in NpO$_2$.\cite{Kubo1,Kubo2,Kubo3}
The definition of multipole as spin-charge density has been
briefly discussed, when we have attempted to clarify
multipole state of heavy lanthanide filled skutterudite.
\cite{Hotta8}
Owing to the definition of multipole in the form of
one-electron spin-charge density operator,
it has been possible to discuss unambiguously
multipole state by evaluating multipole susceptibility
even for heavy rare-earth compounds with
large total angular momentum.

In this paper, first we explain the definition of
multipole as spin-charge density
in the form of one-body operator from the viewpoint of
multipole expansion of electromagnetic potential from
charge distribution in electromagnetism.
In order to determine the multipole state,
we use the optimization of multipole susceptibility
on the basis of the standard linear response theory.
To proceed with further discussion,
in this paper we pick up an impurity Anderson model
including seven $f$ orbitals.
We perform the calculation of multipole susceptibility
by using a numerical renormalization group technique.
Then, we review our recent results of heavy rare-earth
filled skutterudites for Ln=Gd$\sim$Yb.
We also show our new result for multipole susceptibility
of Sm-based filled skutterudites with some comments
on the effect of rattling.
Finally, we summarize this paper and briefly discuss
a future problem of multipole theory.
Throughout this paper,
we use such units as $\hbar$=$k_{\rm B}$=1
and the energy unit is set as eV.

%
%
\section{Multipole as Spin-Charge Density}

In general, multipole is a concept to express the degree of deviation
from spherical symmetric structure.
A well-known example can be found in the multipole expansion of
electromagnetic potential from charge distribution.
The potential is formally expanded by the spherical harmonics $Y_{LM}$
and both electric and magnetic multipole moments are
defined from the expansion coefficients.

Here we emphasize that the multipole expansion is also applicable
for the consideration of electromagnetic field produced by electrons.
\cite{Schwartz}
For instance, Kubo and Kuramoto have discussed
the multipole expansion of the vector potential from local electrons,
in order to estimate the internal magnetic field from octupole moment
in Ce$_{\rm x}$La$_{\rm 1-x}$B$_6$.\cite{Kubo-Kuramoto}

Based on the above background, now we consider the definition of
multipole as $f$-electron density operator ${\hat X}$,
which is generally expressed in the second-quantized form as
\begin{equation}
  {\hat X}=\sum_{m\sigma,m'\sigma'}
  X_{m\sigma,m'\sigma'}f^{\dag}_{m\sigma}f_{m'\sigma'},
\end{equation} 
where $f_{m\sigma}$ is the annihilation operator for $f$ electron with
spin $\sigma$ and $z$-component $m$ of angular momentum $\ell$=3.
Throughout this paper, we define $\sigma$=$1$ ($-1$) for up (down) spin.
The form of the coefficient $X_{m\sigma,m'\sigma'}$ is
related to the definition of multipole.
As easily understood, ${\hat X}$ denotes total charge,
i.e., monopole, for
\begin{equation}
  X^{\rm monopole}_{m\sigma,m'\sigma'}
  =\delta_{mm'}\delta_{\sigma\sigma'}.
\end{equation}
Intuitively, we can understand that $f$-electron multipole denotes
the deviation from the total charge operator.
In the standard definition, dipole denotes total angular momentum $J$.
Thus, ${\hat X}$ for dipole is given by
\begin{equation}
 X^{\rm dipole}_{m\sigma,m'\sigma'}=J^{\alpha}_{m\sigma,m'\sigma'}
 =\delta_{\sigma\sigma'} L^{\alpha}_{mm'}
 +S^{\alpha}_{\sigma\sigma'} \delta_{mm'},
\end{equation}
where $\alpha$ indicates Cartesian component,
$L$ denotes angular momentum operator for $\ell$=3
and $S$ indicates spin operator.
Then, how do we define higher-order multipoles?

In the multipole expansion of potential in electromagnetism,
higher electric and magnetic multipole moments appear
in the coefficients of the expansion by
the spherical harmonics $Y_{LM}$ with larger angular momentum.
In group theory, $Y_{LM}$ is defined by the basis of
irreducible representation $D^{(L)}$ of the rotation group $R$,
expressed as
\begin{equation}
  R Y_{LM}=\sum_{M'} Y_{LM'} D_{MM'}^{(L)}.
\end{equation}
On the analogy of the multipole expansion,
for $f$-electron multipole operator,
we exploit a concept of spherical tensor operator
in the quantum mechanics of angular momentum.\cite{Inui}
When we consider the rotation of operator ${\hat T}$,
we obtain a set of operators ${\hat T}^{(k)}=\{ {\hat T}_q^{(k)} \}$
with $(2k+1)$-components ($q=-k, -k+1, \cdots, k-1, k$), given by
\begin{equation}
  R {\hat T}_{q}^{(k)} R^{-1}
  =\sum_{q'} {\hat T}_{q'}^{(k)}D_{qq'}^{(k)}.
\end{equation} 
Namely, ${\hat T}_q^{(k)}$ is transformed like a basis of
irreducible representation $D^{(k)}$ for the rotation.
Such ${\hat T}_q^{(k)}$ is called spherical tensor operator of rank $k$.

The spherical tensor operator ${\hat T}^{(k)}_q$
for $f$ electron is expressed in the second-quantized form as
\begin{equation}
  {\hat T}^{(k)}_q = \sum_{m\sigma,m'\sigma'}
  T^{(k,q)}_{m\sigma,m'\sigma'}f^{\dag}_{m\sigma}f_{m'\sigma'},
\end{equation}
where the coefficient $T^{(k,q)}_{m\sigma,m'\sigma'}$
is calculated as follows.
First it is convenient to change the $f$-electron basis
from $(m,\sigma)$ to $(j,\mu)$,
where $j$ is the total angular momentum and $\mu$ is
the $z$ component of $j$.
Note that $j$ takes $7/2$ and $5/2$ for $f$ electrons.
For a certain value of angular momentum $j$ and its $z$-component $\mu$,
the matrix element of spherical tensor operator is
easily calculated by the Wigner-Eckart theorem as
\begin{equation}
  \langle j \mu | T_q^{(k)} | j\mu' \rangle
  =\frac{\langle j || T^{(k)} || j \rangle}{\sqrt{2j+1}}
   \langle j\mu | j\mu' kq \rangle,
\end{equation}
where $\langle JM | J'M' J'' M'' \rangle$ denotes
the Clebsch-Gordan coefficient and
$\langle j || T^{(k)} || j \rangle$ is
the reduced matrix element for spherical tensor operator,
given by
\begin{equation}
  \langle j || T^{(k)} || j \rangle=
  \frac{1}{2^k} \sqrt{\frac{(2j+k+1)!}{(2j-k)!}}.
\end{equation}
Note that $k \le 2j$ and the highest rank is $2j$.\cite{note}
The coefficient $T^{(k,q)}_{m\sigma,m'\sigma'}$ is obtained
by returning to the basis of $(m,\sigma)$ from $(j,\mu)$.
The result is given by
\begin{equation}
 \label{Tkq}
 \begin{split}
  T^{(k,q)}_{m\sigma,m'\sigma'}
  &= \sum_{j,\mu,\mu'}
  \frac{\langle j || T^{(k)} || j \rangle}{\sqrt{2j+1}}
  \langle j \mu | j \mu' k q \rangle \\
  &\times 
  \langle j \mu | \ell m s \frac{\sigma}{2} \rangle
  \langle j \mu' | \ell m' s \frac{\sigma'}{2} \rangle,
 \end{split}
\end{equation}
where $\ell$=3, $s$=1/2, $j$=$\ell$$\pm$$s$,
and $\mu$ runs between $-j$ and $j$.

We note that it is not necessary to take double summations
concerning $j$ in eq.~(\ref{Tkq}),
since the matrix representation of total angular momentum
$\mib{J}$ is block-diagonalized in the $(j,\mu)$-basis.
We have checked that the same results as eq.~(\ref{Tkq})
are obtained when we calculate higher-order multipole operators
by following the symmetrized expression of multiple products of
$J$.\cite{Inui,Shiina-multi1,Shiina-multi2}

Thus far, we have implicitly assumed $f$-electron density
in an isolated ion, but in actuality, rare-earth ions are
in the cubic crystal structure.
Then, it is convenient to change from spherical to cubic
tensor operators, given by
\begin{equation}
  {\hat T}^{(k)}_{\gamma}
  =\sum_q G^{(k)}_{\gamma,q}{\hat T}^{(k)}_{q},
\end{equation}
where $k$ is a rank of multipole,
an integer $q$ runs between $-k$ and $k$,
$\gamma$ is a label to express $O_{\rm h}$ irreducible representation,
and $G^{(k)}_{\gamma,q}$ is the transformation matrix
between spherical and cubic harmonics.
Throughout this paper, we use the cubic tensor operator
as multipole.

It should be noted here that multipoles belonging to the same symmetry
are mixed in general, even if the rank is different.
In addition, multipoles are also mixed due to the effect of
crystalline electric field (CEF) of $T_{\rm h}$ point group.
Namely, the $f$-electron spin-charge density should be given by
the appropriate superposition of multipoles,
expressed as
\begin{equation}
  \label{multi}
  {\hat X}=\sum_{k,\gamma} p^{(k)}_{\gamma}{\hat T}^{(k)}_{\gamma}.
\end{equation} 
In order to determine the coefficient $p^{(k)}_{\gamma}$,
we evaluate the multipole susceptibility
in the linear response theory.\cite{Hotta4}
Namely, $p^{(k)}_{\gamma}$ is determined by the normalized eigenstate
of susceptibility matrix, defined as
\begin{equation}
 \begin{split}
  \chi_{k\gamma,k'\gamma'} \!
  = & \frac{1}{Z}
  \sum_{i,j} \frac{e^{-E_i/T}-e^{-E_j/T}}{E_j-E_i}
  \langle i | [{\hat T}^{(k)}_{\gamma}-\rho^{(k)}_{\gamma}] | j \rangle
  \\ & \times 
  \langle j | [{\hat T}^{(k')}_{\gamma'}- \rho^{(k')}_{\gamma'}]| i \rangle,
 \end{split}
\end{equation}
where $E_i$ is the eigenenergy for the $i$-th eigenstate $|i\rangle$ of
the Hamiltonian $H$ of the system, $T$ is a temperature,
$\rho^{(k)}_{\gamma}=\sum_i e^{-E_i/T}
\langle i |{\hat T}^{(k)}_{\gamma}| i \rangle/Z$,
and $Z$ is the partition function given by $Z=\sum_i e^{-E_i/T}$.

When we express the multipole moment as eq.~(\ref{multi}),
we normalize each multipole operator so as to satisfy
the orthonormal condition
Tr$\{ {\hat T}^{(k)}_{\gamma}{\hat T}^{(k')}_{\gamma'} \}$=
$\delta_{kk'}\delta_{\gamma\gamma'}$.\cite{Kubo4}
We note that the multipole susceptibility is given by
the eigenvalue of the susceptibility matrix.
Note also that the susceptibility for 4u multipole moment does
$not$ mean magnetic susceptibility, which is evaluated
by the response of magnetic moment $\mib{L}$+$2\mib{S}$, i.e.,
$\mib{J}$+$\mib{S}$.\cite{Hotta2,Hotta3}

%
%
\section{Model and Parameters}

In the previous section, we have defined multipole
as spin-charge density in the form of one-body operator.
In order to determine the multipole state,
we have explained a method to optimize
the multipole susceptibility based on the linear response theory.
To proceed with further discussion,
it is necessary to set the Hamiltonian $H$.
For $f$-electron systems, it is desirable to treat
a seven-orbital periodic Anderson model, but at least at present,
it seems to be a heavy task to analyze such a multiorbital model.
To reduce the task in calculations by keeping essential physics,
we can take two ways.

One is to decrease the number of relevant $f$ orbitals
in the periodic system.
As for the research along this line, here we briefly introduce
the theoretical study on octupole order in NpO$_2$.
\cite{Kubo1,Kubo2,Kubo3}
After the discussion about the CEF states of actinide dioxides,
two relevant $\Gamma_8$ orbitals have been extracted.
Then, on the basis of a $j$-$j$ coupling scheme,
the $\Gamma_8$ orbital degenerate Hubbard model has been set
as an effective Hamiltonian for NpO$_2$.
With the use of the standard perturbation theory
in terms of $f$-electron hopping,
effective interactions between multipoles
in adjacent sites have been evaluated on an fcc lattice.
Due to the combination of exact diagonalization and mean-field theory,
it has been concluded that the ground state has longitudinal
triple-$\mib{q}$ 5u octupole order,
consistent with experimental facts.

Another way is to keep all seven $f$ orbitals, but to consider
an impurity Anderson model.
Even if the number of local degree of freedom is increased,
it is possible to solve the impurity Anderson model
with the use of a numerical renormalization group method.
In this paper, we adopt this way to discuss multipole state
of filled skutterudites.

The seven-orbital Anderson model for filled skutterudites
is given by
\begin{equation}
  H  \!=\! \sum_{\mib{k},\sigma}
  \varepsilon_{\mib{k}} c_{\mib{k}\sigma}^{\dag} c_{\mib{k}\sigma}
  \!+\! \sum_{\mib{k},\sigma,m}
  (V_{m} c_{\mib{k}\sigma}^{\dag}f_{m\sigma}+{\rm h.c.})
  \!+\! H_{\rm loc},
\end{equation}
where $\varepsilon_{\mib{k}}$ is conduction electron dispersion
and $c_{\mib{k}\sigma}$ is the annihilation operator for conduction
electron with momentum $\mib{k}$ and spin $\sigma$.
The second term indicates
the hybridization between conduction and $f$ electrons.
For filled skutterudites, the main conduction band is given by
$a_{\rm u}$, constructed from $p$-orbitals of pnictogen.\cite{Harima}
Note that the hybridization occurs between the states
with the same symmetry.
Since the $a_{\rm u}$ conduction band has xyz symmetry,
we set $V_2$=$-V_{-2}$=$V$ and zero for other $m$.
We fix $V$ as $V$=0.05 eV and a half of the bandwidth
of $a_{\rm u}$ conduction band is set as 1 eV.

The third term in $H$ is the local $f$-electron part,
given by
\begin{eqnarray}
  H_{\rm loc} = H_{\rm so} + H_{\rm int} + H_{\rm CEF}.
\end{eqnarray}
Note that the chemical potential is appropriately changed
to adjust the local $f$-electron number $n$.
The spin-orbit coupling term $H_{\rm so}$ is given by
\begin{eqnarray}
  H_{\rm so} = \lambda \sum_{m,\sigma,m',\sigma'}
  \zeta_{m,\sigma,m',\sigma'} f_{m\sigma}^{\dag}f_{m'\sigma'},
\end{eqnarray}
where $\lambda$ is the spin-orbit coupling
and $f_{m\sigma}$ denotes the annihilation operator for
$f$ electron with spin $\sigma$ and angular momentum
$m$(=$-3$,$\cdots$,3) and $\sigma$=$+$1 ($-$1) for up (down) spin.
The matrix elements are expressed by
$\zeta_{m,\pm 1,m,\pm 1}$=$\pm m/2$,
$\zeta_{m \pm 1,\mp 1,m, \pm 1}$=$\sqrt{12-m(m \pm 1)}/2$,
and zero for the other cases.
The second term $H_{\rm int}$ indicates the Coulomb interactions
among $f$ electrons, expressed by
\begin{equation}
  H_{\rm int} \!=\! \sum_{m_1 \sim m_4 \atop \sigma_1, \sigma_2}
  I_{m_1,m_2,m_3,m_4}f_{m_1\sigma_1}^{\dag}f_{m_2\sigma_2}^{\dag}
  f_{m_3\sigma_2}f_{m_4\sigma_1},
\end{equation}
where the Coulomb integral $I$ is expressed by the combination of
Slater-Condon parameters, $F^0$, $F^2$, $F^4$, and $F^6$.\cite{Slater}

Finally, the CEF term $H_{\rm CEF}$ is given by
\begin{eqnarray}
  \label{Eq:CEF}
  H_{\rm CEF} = \sum_{m,m',\sigma} B_{m,m'}
  f_{m\sigma}^{\dag}f_{m'\sigma},
\end{eqnarray}
where $B_{m,m'}$ is determined from the table of Hutchings
for angular momentum $J$=$\ell$=3,\cite{Hutchings}
since we are now considering the potential for $f$ electron.
For filled skutterudites with $T_{\rm h}$ symmetry,\cite{Takegahara}
$B_{m,m'}$ is expressed by using three CEF parameters
$B_4^0$, $B_6^0$, and $B_6^2$.\cite{Hutchings}
Following the traditional notation, we define
\begin{eqnarray}
  \begin{array}{l}
    B_4^0=Wx/F(4),\\
    B_6^0=W(1-|x|)/F(6),\\
    B_6^2=Wy/F^t(6),
  \end{array}
\end{eqnarray}
where $x$ and $y$ specify the CEF scheme for $T_{\rm h}$ point group,
\cite{Takegahara}
while $W$ determines an energy scale for the CEF potential.
Concerning $F(4)$, $F(6)$, and $F^t(6)$, we choose
$F(4)$=15, $F(6)$=180, and $F^t(6)$=24 for $\ell$=3.\cite{Hutchings,LLW}

For further discussions, it is necessary to set the parameters.
First let us consider the Slater-Condon parameters $F^k$
($k$=0, 2, 4, and 6) and spin-orbit coupling $\lambda$.
We will discuss CEF parameters later.
In order to determine $F^k$ and $\lambda$,
we try to reproduce the $f^2$ excitation spectrum of Pr$^{3+}$ ion.
\cite{Cai,Eliav}
Here we consider twelve excited states, which are labeled as
$^3H_5$, $^3H_6$, $^3F_2$, $^3F_3$, $^3F_4$, $^1G_4$,
$^1D_2$, $^3P_0$, $^3P_1$, $^3P_2$, $^1I_6$, and $^1S_0$.
By diagonalizing $H_{\rm so}+H_{\rm int}$,
we obtain eigenenergies $E_j$ and
define excitation energy $\delta E_j$=$E_j-E_0$,
where $E_0$ is the ground state energy for $^3H_4$.
We determine $F^k$ and $\lambda$ so as to minimize
the sum of the square of energy difference, given by
$\Delta=\sum_{j=1}^{12} (\delta E_j^{\rm exp}-\delta E_j)^2$,
where $\delta E_j^{\rm exp}$ indicates the excitation energy
experimentally determined.
Note that $F^0$ cannot be determined in the present simple procedure,
since it appears only as the offset of the energy level
when local $f$-electron number is assumed to be unchanged.
In order to determine the value of $F^0$ itself,
it is necessary to resort to the first-principle calculation,
but it is out of the scope of this paper.
Here we simply set by hand $F^0$=10 eV as a typical value
for rare-earth ion.

After the minimization of $\Delta$, we obtain
$F^2$=8.75 eV, $F^4$=6.60 eV, $F^6$=4.44 eV, and $\lambda$=0.095 eV.
These values are expected to be reasonable,
since the experimental value of $\lambda$ is 0.094 eV.
For other lanthanides, we use the same values for the Slater-Condon
parameters, since there is no reason to change two-body interactions,
even when the electron number is changed.
On the other hand, since the spin-orbit interaction is considered to be
sensitive to ion radius, we use experimental values of $\lambda$
for each lanthanide such as $\lambda$=0.144 eV (Sm),
0.180 eV (Gd),  0.212 eV (Tb),
0.240 eV (Dy), 0.265 eV (Ho), 0.295 eV (Er), 0.326 eV (Tm),
and 0.356 eV (Yb).\cite{spin-orbit}

\begin{figure}[t]
\centering
\includegraphics[width=7.0truecm]{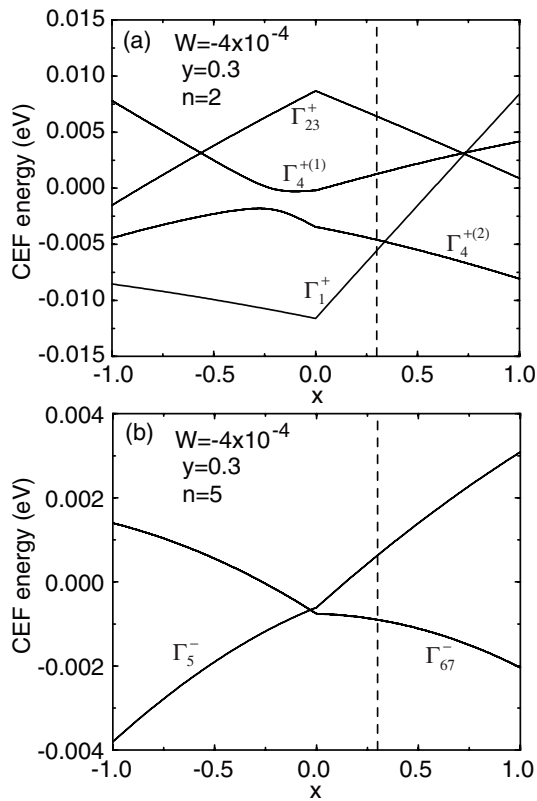}
\caption{CEF energy levels for (a) $n$=2 and (b) $n$=5.}
\end{figure}

Concerning CEF parameters, we emphasize that electrostatic
potentials act on one $f$-electron state.
In principle, it is not necessary to change the CEF parameters
when we substitute rare-earth ion,
as long as we consider the same crystal structure.
Namely, it is enough to determine the CEF parameters
for some filled skutterudite compound.
First we set $W$=$-0.4$ meV and $y$=0.3,
which are considered to be typical values for filled skutterudites.
However, the CEF state is drastically changed by $x$.
From the CEF energy level for $n$=2, as shown in Fig.~1(a),
we choose $x$=0.3 so as to reproduce quasi-quartet CEF scheme
of PrOs$_4$Sb$_{12}$.\cite{Kohgi,Kuwahara,Goremychkin}
We use such CEF parameters for other rare-earth ions.
Since we discuss multipole states for Sm-, Gd-, and Ho-based
filled skutterudites, here we show the CEF energy schemes
for $n$=5, 7, and 10.
Results for other cases are found in Ref.~\citen{Hotta3},
although we used different parameters there.
Namely, the CEF energy schemes are not changed drastically,
as long as we use realistic values for Coulomb interaction
and spin-orbit coupling.

\begin{figure}[t]
\centering
\includegraphics[width=7.0truecm]{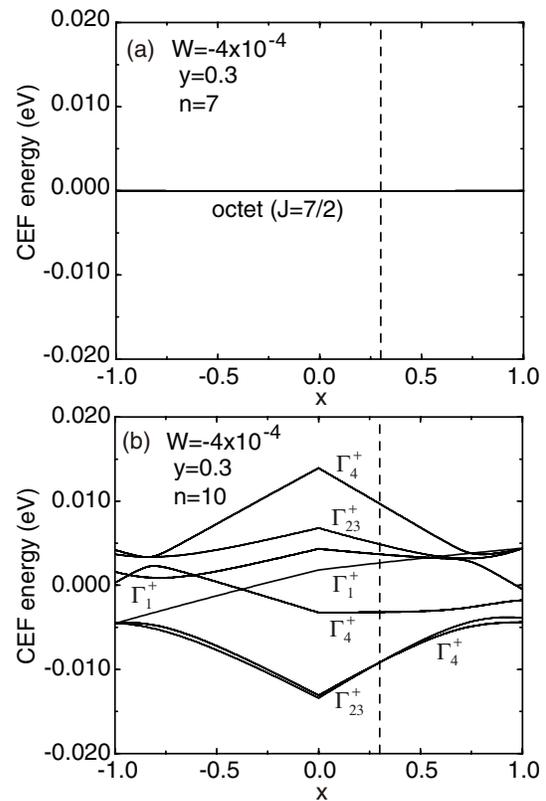}
\caption{CEF energy levels for (a) $n$=7 and (b) $n$=10.}
\end{figure}

In Fig.~1(b), we show the CEF energy levels for $n$=5.
There appear $\Gamma_{5}^-$ doublet and $\Gamma_{67}^-$ quartet
states for $x<0$ and $x>0$, respectively.
At $x$=0.3, we obtain $\Gamma_{67}^-$ quartet ground state.
From the specific heat measurements
for SmRu$_4$P$_{12}$ and SmOs$_4$P$_{12}$,
the CEF ground state has been concluded to be
$\Gamma_{67}^-$ quartet.\cite{Matsuhira}
Recently, magnetization measurement has been performed and
the observed anisotropy has confirmed the $\Gamma_{67}^-$
quartet ground state in SmOs$_4$Sb$_{12}$.\cite{Aoki-Sm}
On the other hand, the CEF energy scheme can be changed
even among the same Sm-based filled skutterudites.
In fact, for SmFe$_4$P$_{12}$, $\Gamma_5^-$ doublet ground state
has been suggested.\cite{Matsuhira,Nakanishi-Sm}
We have explained that such conversion of the CEF ground state
at $n$=5 occurs due to the balance between Coulomb interaction
and spin-orbit coupling.\cite{Hotta7}

In Figs.~2(a) and 2(b), we show the results for $n$=7 and 10,
respectively.
For the case of Gd$^{3+}$ with $n$=7,
the CEF ground state is well described by $L$=0 and $J$=$S$=7/2,
but it is almost independent of $x$,
since the CEF potentials for $L$=0 provide only the energy shift.
Note, however, that the CEF state is given by the mixture of
the $LS$ and $j$-$j$ coupling schemes.
This point will be discussed later again.
For the case of Ho$^{3+}$ with $n$=10,
a remarkable point is that $\Gamma_{23}^+$ doublet and
$\Gamma_4^+$ triplet states are almost degenerate
in the wide range of the values of $x$.
In particular, around at $x$=0.3, the ground state is easily
converted.
Thus, it is necessary to discuss carefully the case of $n$=10.

%
%
\section{Numerical Results}

In order to evaluate the susceptibility matrix, we employ
a numerical renormalization group (NRG) method,\cite{NRG}
in which momentum space is logarithmically discretized
to include efficiently the conduction electrons
near the Fermi energy.
In actual calculations, we introduce a cut-off $\Lambda$ for
the logarithmic discretization of the conduction band.
Due to the limitation of computer resources,
we keep $M$ low-energy states.
In this paper, we set $\Lambda$=5 and $M$=4500.
Note that the temperature $T$ is defined as
$T$=$\Lambda^{-(N-1)/2}$ in the NRG calculation,
where $N$ is the number of the renormalization step.

First let us summarize our recent results for multipole state
of heavy lanthanide filled skutterudites.\cite{Hotta8}
For Ln=Gd, we have found the effect of quadrupole moments
due to the deviation from the $LS$ coupling scheme.
For Ln=Ho, when the CEF ground state is $\Gamma_{23}^+$ doublet,
the exotic state dominated by 2u octupole moment has been observed.
For Ln=Tb and Tm, the CEF ground state is $\Gamma_1^+$ singlet and
we have found no significant multipole moments at low temperatures,
although we cannot exclude a possibility of antiferro 1g+2g ordering.
For Ln=Dy, Er, and Yb, the CEF ground state is $\Gamma_5^-$ doublet,
and the dominant moment is the mixture of 4u and 5u.
Among these cases, we pick up the results of Ln=Gd and Ho,
and review interesting possibilities for the multipole states.

For the case of Gd$^{3+}$ ion, the dominant multipole component is
4u dipole and the secondary components are 3g and 5g quadrupoles.
Note that after the partial screening due to conduction electrons
at extremely low temperatures, quadrupole moments disappear,
but it seems to observe significant contribution from
quadrupole moments.
In the $LS$ coupling scheme, the $f^7$ state is specified by
$J$=$S$=7/2 and $L$=0.
Namely, at the first glance, we do not expect the appearance
of quadrupole moment.
However, we should note that actual situation is always
between the $LS$ and $j$-$j$ coupling schemes.
Namely, some finite contribution of the $j$-$j$ coupling scheme
is included in the ground state.
If very large $\lambda$ is assumed, first $j$=5/2 sextet is
fully occupied and then, one $f$ electron is accommodated
in $j$=7/2 octet.
Thus, this state can be multipole-active.
Recently, it has been observed in GdRu$_4$P$_{12}$ that
$^{101}$Ru NQR frequency exhibits temperature dependence
below a N\'eel temperature $T_{\rm N}$=22K.\cite{Kohori}
This may be interpreted as the effect of quadrupole
due to the deviation from the $LS$ coupling scheme.

For the case of Ho$^{3+}$ ion,
the CEF ground state is $\Gamma_4^+$ triplet at $x$=0.3,
but the first excited state is $\Gamma_{23}^+$ doublet
with very small excitation energy such as $10^{-5}$ eV.
Thus, as shown in Fig.~2(b), Ho-based filled skutterudite
is considered to be in the quasi-quintet situation.
\cite{Hotta3,Yoshizawa2}
For $x$=0.3, we have found several kinds of multipoles,
but the dominant one is always given by the mixture of
4u and 5u from dipole, octupole, dotriacontapole,
and octacosahectapole.

Here we note that the CEF ground state is fragile for $n$=10.
If we slightly decrease $x$, the ground state is easily changed.
Then, we have evaluated multipole susceptibility for $x$=0.25
with $\Gamma_{23}^+$ doublet ground state.
The multipole states at high temperatures are similar to
those for $x$=0.3, but at low temperatures,
we have found 2u multipole state, expressed as
$p^{(3)}_{\rm 2u}$=$0.955$ and $p^{(7)}_{\rm 2u}$=$0.297$.
Namely, the main component is 2u octupole, but there is
contribution of 2u octacosahectapole.
From the elastic constant measurement for HoFe$_4$P$_{12}$,
some anomalous features have been discussed.\cite{Yoshizawa2}
It seems interesting to reexamine experimental results
from the viewpoint of 2u octupole state.

\begin{figure}[t]
\centering
\includegraphics[width=7.0truecm]{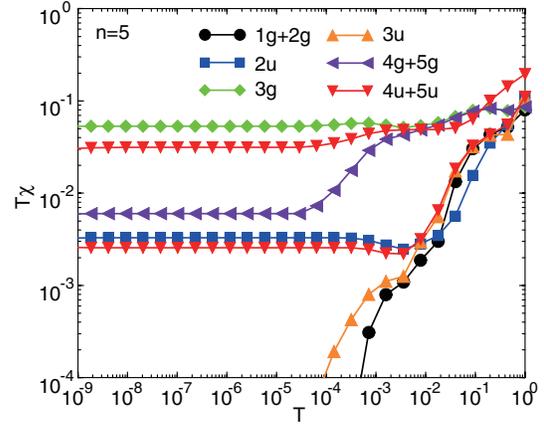}
\caption{(Color online) Multipole susceptibility for $n$=5.}
\end{figure}

Now we show our new result for the case of $n$=5,
corresponding to Sm-based filled skutterudites.
As mentioned in the introduction part,
SmRu$_4$P$_{12}$ has attracted much attention
from a possibility of octupole ordering.
It is meaningful to discuss possible multipole state
in the present calculation.
As shown in Fig.~3, we find that 3g quadrupole is dominant
with $p^{(2)}_{\rm 3g}$=$0.995$ and $p^{(4)}_{\rm 3g}$=$0.097$.
The secondary component is expressed as 4u+5u,
but the contribution from 5u is found to be less than
one percent.
Note that 4u and 5u are mixed due to the $T_{\rm h}$ symmetry,
although such mixing can be included neither in the $LS$ nor
$j$-$j$ coupling scheme.
It is an advantage of the present calculation
to take into account the 4u-5u mixing correctly.
The third and fourth components are, respectively, expressed
by 4g+5g and 2u.
The fifth component is also expressed by 4u+5u,
which includes significant contribution from 5u octupole.

It is difficult to conclude the ordering of multipole
only from the present results,
but the local multipole moment with significant weight
is considered to be a candidate which is ordered in actual system.
Then, if we discuss relevant octupole moment within
the present result of the electronic model,
4u octupole seems to be the best candidate,
since the secondary component includes 4u octupole moment
and the contribution of 5u octupole is very small.

Throughout this paper, we have not included the effect of
phonons, but in the filled skutterudite structure,
anharmonic local phonon, i.e., rattling, has been
considered to play some roles to determine electronic properties.
When we have considered that
the dominant phonon mode in filled skutterudites
is Jahn-Teller type with $E_{\rm g}$ symmetry,\cite{Hotta9}
the multipole state for the case of $n$=5
has been found to be characterized by
the mixture of 4u magnetic and 5u octupole moments
with clear difference between longitudinal and transverse modes.
\cite{Hotta10}
If such symmetry lowering can be detected in experiments,
we should consider seriously the possibility of 5u octupole.
We believe that it is an important test of
5u octupole ordering scenario for the Sm-based filled skutterudite.

%
%
\section{Summary and Comment}

In summary, we have defined the multipole as spin-charge
one-electron density operator by using the cubic tensor operator.
Based on the linear response theory,
we have determined the multipole state so as to maximize
the multipole susceptibility.
Then, we have discussed possible multipole state of
filled skutterudites by evaluating the multipole susceptibility
of the seven orbital Anderson model with the use of NRG method.
After the review of the results for Ln=Gd$\sim$Yb,
we have shown the result for Sm-based filled skutterudites.
Within the impurity Anderson model, quadrupole moment has been
found to be dominant.
We have observed that the secondary component is magnetic
with significant contribution of 4u octupole.
Thus, when we simply ignore the effect of rattling,
4u octupole seems to be relevant in the Sm-based filled skutterudite.
It is consistent with recent result on the specific heat measurement
of SmRu$_4$P$_{12}$.\cite{Aoki-Sm-new}

Finally, we briefly comment on the multipole expansion of $f$-electron
spin-charge density.
In this paper, the spin-charge density has been expressed
by the linear combination of cubic tensor operators
in the cubic crystal structure.
Unfortunately, it does not provide the general multipole
expansion of $f$-electron spin-charge density,
since we are restricted only in the cubic symmetry.
It is one of future issues to complete the multipole expansion
of $f$-electron spin-charge density.

%
%
\section*{Acknowledgment}

The author thanks K. Kubo for valuable comments and useful discussions.
He also thanks Y. Aoki for discussion.
This work has been supported by a Grant-in-Aid for Scientific Research
in Priority Area ``Skutterudites'' under the contract No.~18027016
from the Ministry of Education, Culture, Sports, Science, and
Technology of Japan.
The author has been also supported by a Grant-in-Aid for
Scientific Research (C) under the contract No.~18540361
from Japan Society for the Promotion of Science.
The computation in this work has been done using the facilities
of the Supercomputer Center of Institute for Solid State Physics,
University of Tokyo.

%
%


\begin{thebibliography}{99}

\bibitem{Kuramoto}
Y. Kuramoto:
JPSJ Online-News and Comments [Apr. 13, 2007].

\bibitem{Kusunose}
H. Kusunose:
JPSJ Online-News and Comments [Sep. 10, 2007].

\bibitem{Takegahara}
K. Takegahara, H. Harima and A. Yanase:
J. Phys. Soc. Jpn. {\bf 70} (2001) 1190;
{\it ibid.} {\bf 70} (2001) 3468;
{\it ibid.} {\bf 71} (2002) 372.

\bibitem{Aoki}
Y. Aoki, T. Namiki, S. Ohsaki, S. R. Saha, H. Sugawara and H. Sato:
J. Phys. Soc. Jpn. {\bf 71} (2002) 2098.

\bibitem{Tayama}
T. Tayama, T. Sakakibara, H. Sugawara, Y. Aoki and H. Sato:
J. Phys. Soc. Jpn. {\bf 72} (2003) 1516.

\bibitem{Shiina}
R. Shiina and Y. Aoki:
J. Phys. Soc. Jpn. {\bf 73} (2004) 541.

\bibitem{Takimoto}
T. Takimoto:
J. Phys. Soc. Jpn. {\bf 75} (2006) 034714.

\bibitem{Iwasa}
K. Iwasa, L. Hao, K. Kuwahara, M. Kohgi, S. R. Saha, H. Sugawara,
Y. Aoki, H. Sato, T. Tayama and T. Sakakibara:
Phys. Rev. B {\bf 72} (2005) 024414,

\bibitem{Kiss}
A. Kiss and Y. Kuramoto:
J. Phys. Soc. Jpn. {\bf 75} (2006) 103704.

\bibitem{Sakai}
O. Sakai, J. Kikuchi, R. Shiina, H. Sato, H. Sugawara,
M. Takigawa and H. Shiba:
J. Phys. Soc. Jpn. {\bf 76} (2007) 024710.

\bibitem{Kikuchi}
J. Kikuchi, M. Takigawa, H. Sugawara and H. Sato:
J. Phys. Soc. Jpn. {\bf 76} (2007) 043795.

\bibitem{Nakanishi}
Y. Nakanishi, T. Kumagai, M. Yoshizawa, H. Sugawara and H. Sato:
Phys. Rev. B {\bf 69} (2004) 064409.

\bibitem{Yoshizawa}
M. Yoshizawa, Y. Nakanishi, M. Oikawa, C. Sekine, I. Shirotani,
S. R. Saha, H. Sugawara and H. Sato:
J. Phys. Soc. Jpn. {\bf 74} (2005) 2141.

\bibitem{Hachitani}
K. Hachitani, H. Fukazawa, Y. Kohori, I. Watanabe, C. Sekine
and I. Shirotani:
Phys. Rev. B {\bf 73} (2006) 052408.

\bibitem{Masaki1}
S. Masaki, T. Mito, N. Oki, S. Wada and N. Takeda:
J. Phys. Soc. Jpn. {\bf 75} (2006) 053708.

\bibitem{Masaki2}
S. Masaki, T. Mito, M. Takemura, S. Wada, H. Harima, D. Kikuchi,
H. Sato, H. Sugawara, N. Takeda and G.-q. Zheng:
J. Phys. Soc. Jpn. {\bf 76} (2007) 043714.

\bibitem{Aoki-Sm-new}
Y. Aoki, S. Sanada, D. Kikuchi, H. Sugawara and H. Sato:
J. Phys. Soc. Jpn. {\bf 76} (2007) 113703.

\bibitem{Hotta1}
T. Hotta: Rep. Prog. Phys. {\bf 69} (2006) 2061.

\bibitem{Kubo1}
K. Kubo and T. Hotta:
Phys. Rev. B {\bf 71} (2005) 140404(R).

\bibitem{Kubo2}
K. Kubo and T. Hotta:
Phys. Rev. B {\bf 72} (2005) 132411.

\bibitem{Kubo3}
K. Kubo and T. Hotta:
Phys. Rev. B {\bf 72} (2005) 144401.

\bibitem{Hotta2}
T. Hotta:
Phys. Rev. Lett. {\bf 94} (2005) 067003.

\bibitem{Hotta3}
T. Hotta:
J. Phys. Soc. Jpn. {\bf 74} (2005) 1275.

\bibitem{Hotta4}
T. Hotta:
J. Phys. Soc. Jpn. {\bf 74} (2005) 2425.

\bibitem{Hotta5}
T. Hotta and H. Harima:
J. Phys. Soc. Jpn. {\bf 75} (2006) 124711.

\bibitem{Hotta6}
T. Hotta:
J. Magn. Magn. Mater. {\bf 310} (2007) 1691.

\bibitem{Hotta7}
T. Hotta:
J. Phys. Soc. Jpn. {\bf 76} (2007) 034713.

\bibitem{Hotta8}
T. Hotta:
J. Phys. Soc. Jpn. {\bf 76} (2007) 083705.

\bibitem{Schwartz}
C. Schwartz: Phys. Rev. {\bf 97} (1955) 380.

\bibitem{Kubo-Kuramoto}
K. Kubo and Y. Kuramoto:
J. Phys. Soc. Jpn. {\bf 73} (2004) 216.

\bibitem{Inui}
T. Inui, Y. Tanabe and Y. Onodera:
{\it Group Theory and Its Applications in Physics},
(Springer, Berlin, 1996).

\bibitem{note}
When we define multipoles as tensor operators in the space of
total angular momentum $J$ on the basis of the $LS$ coupling scheme,
there appear multipoles with $k \ge 8$ for the cases of $J \ge 4$,
i.e., for $2 \le n \le 4$ and $8 \le n \le 12$,
where $n$ is local $f$-electron number.
If we need such higher-rank multipoles with $k \ge 8$,
it is necessary to consider many-body operators
beyond the present one-body definition.

\bibitem{Shiina-multi1}
R. Shiina, H. Shiba and P. Thalmeier:
J. Phys. Soc. Jpn. {\bf 66} (1997) 1741.

\bibitem{Shiina-multi2}
R. Shiina: J. Phys. Soc. Jpn. {\bf 73} (2004) 2257.

\bibitem{Kubo4}
K. Kubo and T. Hotta:
J. Phys. Soc. Jpn. {\bf 75} (2006) 013702.

\bibitem{Harima}
H. Harima and K. Takegahara:
J. Phys.: Condens. Matter {\bf 15} (2002) S2081.

\bibitem{Slater}
J. C. Slater:
{\it Quantum Theory of Atomic Structure},
(McGraw-Hill, New York, 1960).

\bibitem{Hutchings}
M. T. Hutchings:
Solid State Phys. {\bf 16} (1964) 227.

\bibitem{LLW}
K. R. Lea, M. J. M. Leask and W. P. Wolf:
J. Phys. Chem. Solids {\bf 23} (1962) 1381.

\bibitem{Cai}
Z. Cai, V. Meiser and C. F. Fischer:
Phys. Rev. Lett. {\bf 68} (1992) 297.

\bibitem{Eliav}
E. Eliav, U. Kaldor and Y. Ishikawa:
Phys. Rev. A {\bf 51} (1995) 225.

\bibitem{spin-orbit}
S. H\"ufner:
{\it Optical Spectra of Transparent Rare Earth Compounds},
(Academic Press, New York, 1978).

\bibitem{Kohgi}
M. Kohgi, K. Iwasa, M. Nakajima, N. Metoki, S. Araki, N. Bernhoeft,
J.-M. Mignot, A. Gukasov, H. Sato, Y. Aoki and H. Sugawara:
J. Phys. Soc. Jpn. {\bf 72} (2003) 1002.

\bibitem{Kuwahara}
K. Kuwahara, K. Iwasa, M. Kohgi, K. Kaneko, S. Araki, N. Metoki,
H. Sugawara, Y. Aoki and H. Sato:
J. Phys. Soc. Jpn. {\bf 73} (2004) 1438.

\bibitem{Goremychkin}
E. A. Goremychkin, R. Osborn, E. D. Bauer, M. B. Maple,
N. A. Frederick, W. M. Yuhasz, F. M. Woodward and J. W. Lynn:
Phys. Rev. Lett. {\bf 93} (2004) 157003.

\bibitem{Matsuhira}
K. Matsuhira, Y. Doi, M. Wakeshima, Y. Hinatsu, H. Amitsuka,
Y. Shimaya, R. Giri, C. Sekine and I. Shirotani:
J. Phys. Soc. Jpn. {\bf 74} (2005) 1030.

\bibitem{Aoki-Sm}
Y. Aoki, S. Sanada, H. Aoki, D. Kikuchi, H. Sugawara and H. Sato:
Physica B {\bf 378-380} (2006) 54.

\bibitem{Nakanishi-Sm}
Y. Nakanishi, T. Tanizawa, T. Fujino, H. Sugawara, D. Kikuchi, H. Sato
and M. Yoshizawa:
{\it Proc. 5th Int. Symp. ASR-WYP-2005: Advances in the Physics and
Chemistry of Actinide Compounds},
J. Phys. Soc. Jpn. {\bf 75} (2006) Suppl., p. 192.

\bibitem{NRG}
H. R. Krishna-murthy, J. W. Wilkins and K. G. Wilson:
Phys. Rev. B {\bf 21} (1980) 1003.

\bibitem{Kohori}
Y. Kohori: private communications.

\bibitem{Yoshizawa2}
M. Yoshizawa, Y. Nakanishi, T. Fujino, P. Sun, C. Sekine and I. Shirotani:
J. Magn. Magn. Mater. {\bf 310} (2007) 1786.

\bibitem{Hotta9}
T. Hotta:
Phys. Rev. Lett. {\bf 96} (2006) 197201.

\bibitem{Hotta10}
T. Hotta:
J. Phys. Soc. Jpn. {\bf 76} (2007) 034713.

\end{thebibliography}
\end{document}